\documentclass[manuscript]{aastex}
\usepackage{spr-astr-addons}
\usepackage{url}
\usepackage{hyperref}
\hypersetup{colorlinks,citecolor=blue,linkcolor=blue,urlcolor=blue}

\slugcomment{Not to appear in Nonlearned J., 45.}

\shorttitle{Light scattering properties of aggregates}
\shortauthors{Deb Roy et al.}

\begin{document}


\title{Study of light scattering properties of dust aggregates with a wide variation of porosity}


\author{P. Deb Roy}
\affil{Department of Physics, Assam University, Silchar 788011, India.}

\and
\author{P. Halder}
\affil{Department of Physics, Assam University, Silchar 788011, India.}

\author{H. S. Das}
\affil{Department of Physics, Assam University, Silchar 788011, India.}




\begin{abstract}
We study the light scattering properties of moderately large dust aggregates ($0.8\mu m \lesssim R \lesssim 2.0\mu m$) with a wide variation of porosity ($\mathcal{P}$) from 0.57 to 0.98. The computations are performed using the Superposition T-matrix code with BAM2 cluster ($\mathcal{P} \sim 0.57-0.64$), BAM1 cluster ($\mathcal{P} \sim 0.74$), BA or BPCA cluster ($\mathcal{P} \sim 0.85-0.87$) and BCCA cluster ($\mathcal{P} \sim 0.98$). The simulations are executed at two wavelengths 0.45$\mu m$ and 0.65$\mu m$ with highly absorbing particles (organic refractory) as well as with low absorbing particles (amorphous silicates) to understand the photopolarimetric behavior (phase function, polarization, and color) of dust aggregates. The effect of aggregate size parameter ($X$) on the light scattering properties of aggregates (BA and BAM2) having different porosities is explored in this study. We find that the positive polarization maximum ($P_{max}$), the amplitude of the negative polarization ($P_{min}$) and phase function at the exact backscattering direction ($S_{11}$(180$^\circ$)) are correlated with the porosity of aggregates. Compact aggregates show deeper negative polarization as compared to porous aggregates when the characteristic radius ($R$) of the aggregates are considered to be the same. Further lower porosity aggregates show higher $S_{11}$(180$^\circ$) and vice versa. When $\mathcal{P}$ is increased in a range from 0.64 to 0.98, both ${S}_{11}(180^\circ$) and $P_{min}$ decrease linearly, whereas P$_{max}$ increases linearly. We also find that the porosity of the aggregates plays a crucial role in determining the polarimetric color for high absorbing organic refractories. The compact clusters (BAM1 and BAM2) show the negative polarimetric color whereas BA clusters show almost positive polarimetric color at all values of scattering angle. We have also made some comparisons of our simulated results with PROGRA$^2$ experimental results.
\end{abstract}





\section{Introduction}
Cosmic dust is found to be in the form of aggregates of small constituent particles which includes asteroidal dust, comet dust, interstellar dust, and interplanetary dust. The intensity and linear polarization of the scattered light are the two main source of information regarding the optical and physical properties of the dust aggregates. Comet is found to be a good polarizer of light due to the scattering of sunlight from the cometary dust and the fluorescence emission by the gaseous molecules. The observed polarization in comet depends on the size and shape of the dust particles, the wavelength of the incident light, refractive indices of the dust particles and more importantly the Sun- Comet- Earth phase angle. The phase functions of brightness and polarization for the cometary dust and the regolith on most of the atmosphereless solar system bodies show qualitatively similar feature \citep{Ko2004}. The dependence of linear polarization on the scattering angle shows a bell-shaped curve with its maximum value of typically $10 - 30\%$ at the scattering angle range between $80^\circ$ to $90^\circ$ \citep{Le1996}. A pronounced negative polarization branch observed at high scattering angle in the phase curve where the transition from positive to negative polarization takes place at the scattering angle $160^\circ$ with a minimum of approximately $- 2\%$ \citep{Ea1992,Do1988}.

The local polarimetric properties of dust particles are different in different regions in the cometary coma \citep{Re1996}. The non-uniformity in the distribution of polarization can also be noticed in the special structures like jets, fans, etc. To explain these observed photopolarimetric features of comets, numerical simulations of light scattering characteristics have been proposed by many researchers to construct `a plausible model of comet dust' \citep{Ko2004, Da2004, Da2006, Da2008, Da2011, Ki2016}. Recently, \cite{Maz2017} have investigated the correlation among different scattering parameters e.g., the positive polarization maximum, the amplitude of negative polarization, complex refractive indices, geometric albedo etc. to study the light scattering properties of aggregate particles in a wide range of complex refractive indices and wavelengths.

The intensity in the forward-scattering domain where all the scattered light rays are in phase depends only on the size of the aggregate itself regardless of the properties of constituent particles. But at the high scattering angle, the scattered intensity depends on the degree of coherence of the scattered light which in turn related to the individual monomers \citep{Mu1991}. Thus the degree of polarization at the high scattering angles is sensitive to the properties of the constituent monomers of the aggregate such as the size, the absorption index and the porosity of the aggregate.  In this regard, we emphasize primarily on the backscattering region to retrieve the photopolarimetric properties of the cosmic dust aggregates at the high scattering angle.
					
The main objective of the paper is to study the light scattering properties of moderately large dust aggregates ($0.8\mu m \lesssim R \lesssim 2.0\mu m$) as compared to the wavelength of radiation with a wide variation of porosity from 0.57 to 0.98. We also scrutinize the effect of composition, the effect of the size of the aggregates and porosity especially on the negative branch of polarization observed in the near-nucleus region of comets when the scattering angle is higher than 160$^\circ$ in different sections of the paper. The correlation between the enhancement in intensity and the production of deeper and wider negative polarization branch (NPB) for the moderately large dust aggregates of wide porosity range at the backscattering regime are also of a major interest in this study.

The structure of the paper is as follows. Firstly, we describe the type of considered large aggregates used for model calculations to investigate the light scattering properties of dust aggregates. Secondly, we emphasize on the effect of material composition, the characteristic radius, and porosity of the aggregate on the photopolarimetric properties of dust aggregates. Finally, we systematically discuss the effect of mixing on the photopolarimetric properties of dust aggregates.

\section{Aggregate Dust Model}
To study the natural aggregates, several investigators built the aggregates using the ballistic aggregation procedure. The aggregates have been built using the Monte-Carlo simulation by random hitting and sticking particles together. If the procedure permits a single particle to join a cluster of particles, the aggregate is called Ballistic Particle-Cluster Aggregate (BPCA) and if the procedure allows clusters of particles to stick together then the aggregate is called Ballistic Cluster-Cluster Aggregate (BCCA).  Later, \cite{Sh2008}  introduced three different models of cluster growth, such as Ballistic Aggregates (BA), Ballistic Agglomeration with one migration (BAM1) and Ballistic Agglomeration with two migration (BAM2).  BA structure is generated by Ballistic Agglomeration process having porosity $\approx 0.87$ and fractal dimension $\approx 3.0$, which is similar to the BPCA.  The geometries of BAM1 \& BAM2 clusters are random but substantially less ``fluffy'' than the BA clusters with porosity $\approx 0.74$ and $\approx 0.64$ respectively. The structures of BA, BAM1, and BAM2 clusters are depicted in Fig. 1. The porosity or the openness of the cluster is defined as $\mathcal{P} = 1-(a_{eff}$/R)$^3$, where $a_{eff}$ is the radius of the equi-volume sphere radius, and R is the characteristic radius of the aggregate. \cite{Sh2008} calculated the characteristic radius of the aggregate by comparing the particles to an equivalent ellipsoid obtained from fitting the moment of inertia of the particle. Then a geometric mean of the three ellipsoid axes is taken to obtain the characteristic radius ($R$) of the aggregate.

In this paper, we are primarily interested to study the effect of porosity on the light scattering properties of dust aggregates. So we adopt BCCA, BA, BAM1, and BAM2 clusters as they provide a wide range of porosity from very fluffy BCCA ($\mathcal{P}$$\sim$0.98), BA ($\mathcal{P}$$\sim$0.87) to moderately porous BAM1 ($\mathcal{P}$$\sim$0.74) and very compact BAM2 clusters ($\mathcal{P}$$\sim$0.57). The specified structures (BA, BAM1 and BAM2) for our calculation purpose are taken from  Bruce T. Draine's website\footnote{https://astro.princeton.edu/~draine/agglom.html}. The prescriptions for producing random aggregates of spheres are discussed in \cite{Sh2008}.  The structural parameters of the aggregates used in our computation purpose are shown in Table-1.

\begin{figure*}\label{Fig1}
\begin{center}
\includegraphics[width=120mm]{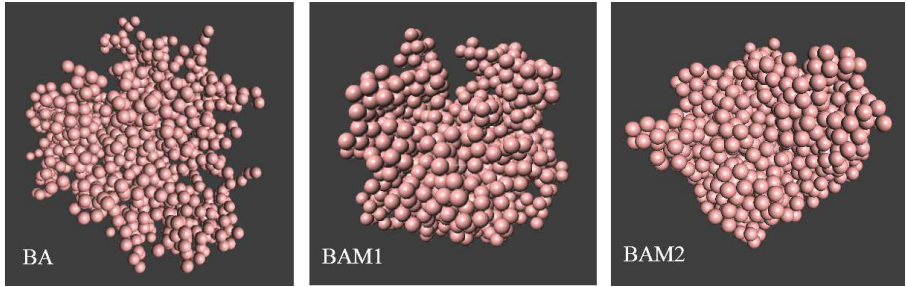}

	  \caption{BA, BAM1 and BAM2 clusters with 1024 monomers having porosities $\sim 0.87$, 0.74 and 0.64 respectively.}
\end{center}

\end{figure*}

\begin{table*}
\vspace{0.1cm}
\caption{Cluster geometries from \cite{Sh2008}: Types of aggregate, number of monomers ($N$), structure number as mentioned in \cite{Sh2008}, porosity of the aggregate ($\mathcal{P}$), ratio of aggregate radius to the effective equal volume radius ($R/a_{eff}$), monomer radius ($a_m$), effective equal volume radius ($a_{eff}$), characteristic radius of the aggregate ($R$) and the size parameter of the aggregate ($X = 2 \pi R/\lambda$) at two wavelengths 0.45$\mu$m and 0.65$\mu$m. }
\begin{center}
\begin{tabular}{|c|c|c|c|c|c|c|c|c|c|}
\hline
   Aggregate & $N$ & Structure & $\mathcal{P}$ & $R/a_{eff}$  & $a_m$ & $a_{eff}$  & $R$ & $X$ \\
   &       &                  &                      &           &($\mu$m) &($\mu$m) & ($\mu$m)    & 0.45$\mu$m \vline 0.65$\mu$m \\
 \hline
   BA &  256 & 1,10,15 & 0.85 & 1.9 & 0.1 & 0.63 & 1.19& 16.6 \vline 11.5\\

    &  512 & 1,2,3 & 0.86 & 1.93 & 0.1 & 0.8 & 1.54& 21.5 \vline 14.9\\

    &  1024 & 1,2,3 & 0.87 & 1.95 & 0.1 & 1.0 & 1.95& 27.3 \vline 18.9\\

    &  1024 & 1,2,3 & 0.87 & 1.95 & 0.05 & 0.5 & 1.0& 13.9 \vline 9.67\\
    \hline
     BAM1 & 1024 & 1,10,11 & 0.74 & 1.58 & 0.1 & 1.0 & 1.58 & 21.5 \vline 15.3\\

    &  1024 & 1,10,11 & 0.74 & 1.58 & 0.06 & 0.63 & 1.0 & 13.9 \vline 9.67\\
   \hline
   BAM2 & 256 & 1,5,7 & 0.57 & 1.33 & 0.1 & 0.63 & 0.84 & 11.8 \vline 8.12\\
        & 512 & 1,2,3 & 0.61 & 1.37 & 0.1 & 0.8  & 1.09 & 15.2 \vline 10.5\\

    &  1024 & 1,2,3 & 0.64 & 1.41 & 0.1 & 1.0 & 1.41 & 19.7 \vline 13.6\\
    &  1024 & 1,2,3 & 0.64 & 1.41 & 0.07 & 0.7 & 1.0 & 13.9 \vline 9.67\\
   \hline
\end{tabular}
\end{center}
\end{table*}


The simulations are executed at two wavelengths 0.45$\mu m$ and 0.65$\mu m$ with highly absorbing (organic refractory) as well as with low absorbing particles (amorphous silicates) to understand the photopolarimetric behavior (phase function, polarization, and color) of aggregated dust particles.
	
It is well known from \emph{in situ} measurement on comet Halley that the cometary dust contains mainly of magnesium-rich silicates, carbonaceous materials and iron bearing sulfides \citep{Je1988, Je1999}. These materials are the major constituents of interplanetary dust particles (IDPs) \citep{Br1980}. Organic materials, amorphous and crystalline silicate minerals (e.g. forsterite, enstatite) are also detected in comets and IDPs \citep{Ha2004}. The infrared (IR) measurement of comets has shown the predominance of both  amorphous and crystalline silicates consisting of pyroxene or olivine grains \citep{Wo1999,Ha2000,Bo2002}. \emph{Stardust} samples confirmed the presence of a variety of olivine and pyroxene silicates in Comet 81P/Wild 2 \citep{Zo2006} including organics \citep{Sa2006}. The amorphous magnesium silicates, e.g., amorphous silicate with stoichiometry forsterite composition are the main component of cometary silicates \citep{Ko2015, Ha2004}. This is consistent with the \emph{Stardust} returned samples, IR spectra, \emph{in situ} Giotto data, and interplanetary dust particles \citep{Ri2008}. Recently, the results obtained from the Rosetta mission on comet 67P/Churyumov-Gerasimenko reveal the presence of large amount of organic materials \citep{Sc2015, Ca2015, Go2015}. With this background, we study the light scattering properties of cosmic dust aggregates by considering the amorphous silicate and the organic refractory composites.  We consider the wavelength dependent refractive indices in our work.  The refractive indices of the amorphous silicates are $m = 1.689 + i~ 0.0031$ at $\lambda$= 0.45$\mu$m and $m = 1.677 + i~ 0.0044$ at $\lambda$ = 0.65$\mu$m \citep{Sc1996}. A few computation has also been performed for the organic refractory of refractive indices $m = 1.813 + i~ 0.479$ at $\lambda$= 0.45$\mu$m and $m = 1.93 + i~ 0.367$ at $\lambda$=0.65$\mu$m \citep{Je1993}. This set of refractive indices was also used by \cite{Ko2015}.
 	
In our simulations, we have considered three cluster realizations of a particular type of aggregates (BA, BAM1 or BAM2) having the same porosity ($\mathcal{P}$) and R/a$_{eff}$ to reduce the variation in the porosity of the aggregates (see Table-1). The porosity along with its error for BA, BAM1 and BAM2 clusters with N = 256, 512  and 1024 are given in \cite{Sh2008} (see Table-2 of that paper).  We have also considered highly porous BCCA cluster in our computations. We perform our calculations using parallel Multi-sphere T-matrix code version 3.0 developed by \cite{Ma2013}. This code is highly efficient in the parallel computational environment and could be used to study the light scattering properties of large aggregates.

 \section{Results and Discussion}

\subsection{Effect of composition}

We now study the effect of composition on the polarization of the scattered light and the phase function ($S_{11}$) for moderately large aggregates compared to wavelength of incident radiation. This dependence was already studied by many investigators for small aggregates in the past \citep{Pe2004, Ki2006, Be2007, Ma2015}. It is to be noted that the phase function is directly related to the albedo of the particles. The geometric albedo ($A$) can be calculated using the formula defined by Hanner et al. (1981), which is given by $[{S}_{11}(180^\circ)~\lambda ^2]/(4\pi G)$. Here ${S}_{11}(180^\circ)$ is the first element of the scattering matrix at the exact back scattering direction and $G$ is the geometric cross section of aggregates. In Figs. 2 and 3, the angular dependence of phase function ($S_{11}$) and polarization ($-S_{12}/S_{11}$) of BA, BAM1 and BAM2 clusters with 1024 numbers of constituent monomers are presented for the amorphous silicate (MgSiO$_2$) with the complex refractive index $m = 1.689 + i~ 0.0031$ at $\lambda$= 0.45$\mu$m and $m = 1.677 + i~ 0.0044$ at $\lambda$ = 0.65$\mu$m, and organic refractory composites with $m = 1.813 + i~ 0.479$ at $\lambda$= 0.45$\mu$m and $m = 1.93 + i~ 0.367$ at $\lambda$=0.65$\mu$m.  It is to be noted that the wavelength dependent refractive indices are considered in this computations. The porosity ($\mathcal{P}$) and the characteristic radius ($R$) for BA, BAM1 and BAM2 clusters are (0.87, 1.95$\mu$m), (0.74, 1.58$\mu$m) and (0.64, 1.41$\mu$m) respectively. These are taken to be the same at the two wavelengths 0.45$\mu$m and 0.65$\mu$m (see Table-1).

\begin{figure*}\label{Fig2}
\begin{center}
\includegraphics[width=120mm]{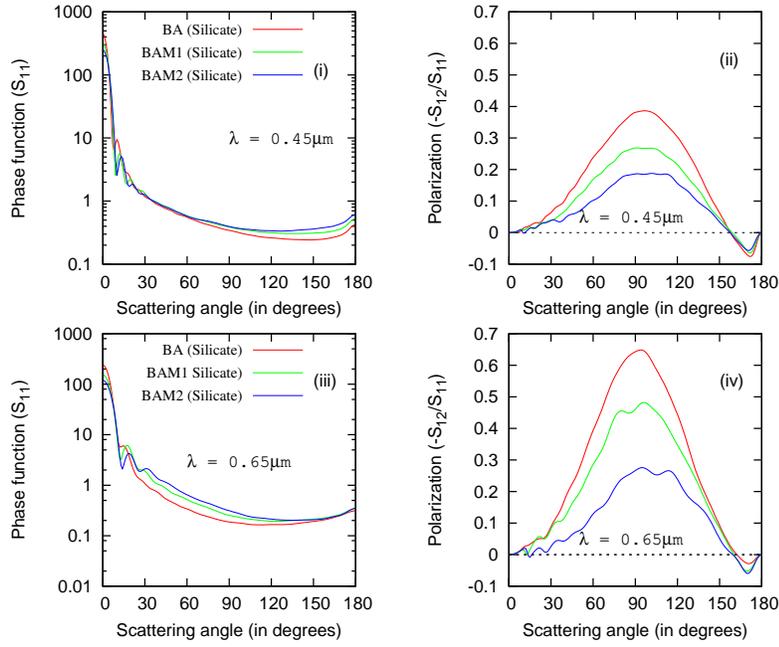}

	  \caption{Photopolarimetric characteristics of BA, BAM1 and BAM2 clusters consisting of amorphous silicate composites with the complex refractive index $m = 1.689 + i~ 0.0031$ at $\lambda$= 0.45$\mu$m and $m = 1.677 + i~ 0.0044$ at $\lambda$ = 0.65$\mu$m. The characteristic radius (R$\sim$1.95$\mu$m, 1.6$\mu$m, and 1.4$\mu$m) and porosity ($\mathcal{P}$$\sim$0.87, 0.74 and 0.64) of BA, BAM1 and BAM2 clusters are taken to be the same at the two wavelengths 0.45$\mu$m and 0.65$\mu$m. Phase function and polarization of the aggregate obtained from this work at $\lambda$=0.45$\mu$m (i and ii) and 0.65$\mu$m (iii and iv) are shown.}
\end{center}

\end{figure*}

\begin{figure*}
\begin{center}
\includegraphics[width=120mm]{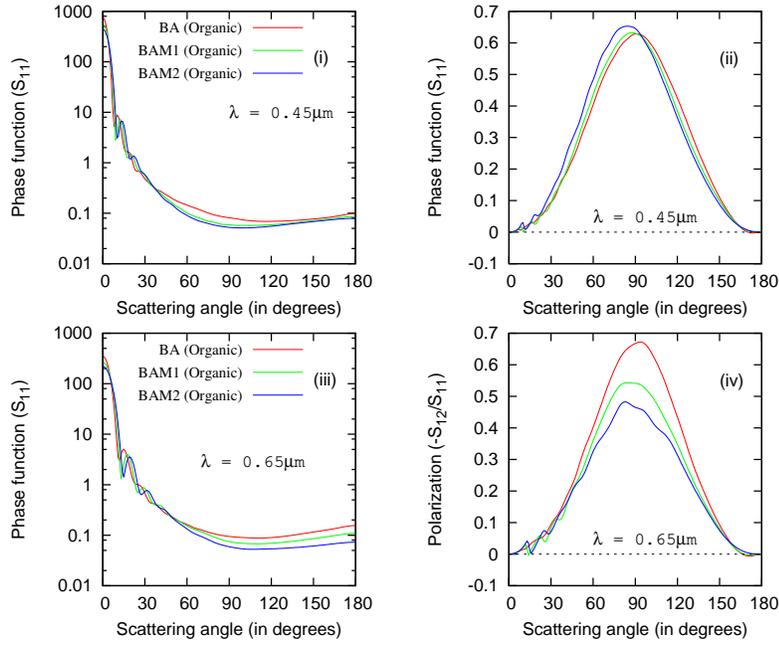}

  \caption{Photopolarimetric characteristics of BA, BAM1 and BAM2 clusters consisting of organic refractory composites with $m = 1.813 + i~ 0.479$ at $\lambda$= 0.45$\mu$m and $m = 1.93 + i~ 0.367$ at $\lambda$=0.65$\mu$m. The characteristic radius (R$\sim$1.95$\mu$m, 1.6$\mu$m and 1.4$\mu$m) and porosity ($\mathcal{P}$$\sim$0.87, 0.74 and 0.64) of BA, BAM1 and BAM2 clusters are taken to be the same at the two wavelengths 0.45$\mu$m and 0.65$\mu$m. Phase function and polarization of the aggregate obtained from this work at $\lambda$=0.45$\mu$m (i and ii) and 0.65$\mu$m (iii and iv) are shown.}
\end{center}

\end{figure*}


Figs. 2 and 3 depict the increase in the degree of polarization with the increase in wavelength of the BA, BAM1 and BAM2 clusters for amorphous silicate composition. But surprisingly, a decrease in polarization maximum for the organic refractory composition is being well observed with the increase in wavelength of radiation for compact BAM1 and BAM2 clusters. The scattering angle of polarization maximum ($\theta_{max}$) values for BA, BAM1 and BAM2 clusters in case of silicate compositions are given by $97^\circ$ (at $\lambda = 0.45 \mu m$) and $94^\circ$ (at $\lambda = 0.65 \mu m$), $91^\circ$ (at $\lambda = 0.45 \mu m$) and $96^\circ$ (at $\lambda = 0.65 \mu m$), and $91^\circ$ (at $\lambda = 0.45 \mu m$) and $95^\circ$ (at $\lambda = 0.65 \mu m$), respectively. In case of organic refractories, the values are $91^\circ$ (at $\lambda = 0.45 \mu m$) and $94^\circ$ (at $\lambda = 0.65 \mu m$), $88^\circ$ (at $\lambda = 0.45 \mu m$) and $84^\circ$ (at $\lambda = 0.65 \mu m$), and $84^\circ$ (at $\lambda = 0.45 \mu m$) and $83^\circ$ (at $\lambda = 0.65 \mu m$), respectively. The scattering angle of polarization maximum ($\theta_{max}$) for all three clusters attains at a higher scattering angle for the amorphous silicate compositions as compared to the case of organic refractories.  In contrast to the high negative polarization observed for amorphous silicate ($-7.6\%$ for BA and $-5.6\%$ for BAM2 at $\lambda$=0.45$\mu$m; and $-2.9\%$ and $-6.0\%$ at $\lambda$=0.65$\mu$m), the aggregates of organic refractory produce a small negative polarization ($-0.3\%$ and $-0.6\%$ for BA) at high scattering angle. But no negative polarization is being noticed for BAM2 with organic composition. It is also important to notice that Figure 2 shows a strong branch of negative polarization with the high albedo amorphous silicate composition irrespective of the variation in wavelength of radiation as well as the nature of the aggregates.
	
The left panel of Figs. 2 and 3 show the phase function of the BA, BAM1 and BAM2 clusters for the composition of amorphous silicates and organic refractory at $\lambda$=0.45 and 0.65$\mu$m. The silicate composites show high scattering efficiency, which results in a higher scattering phase function at 180$^\circ$ scattering angle (${S}_{11}(180^\circ$)) for BA, BAM1 and BAM2 clusters. More interestingly, the aggregates of silicate composition show an increment in ${S}_{11}(180^\circ$) with the decrease in wavelength whereas the highly absorbing organic refractory depicts an opposite trend and possess higher ${S}_{11}(180^\circ$) at the higher wavelength. This may be due to combined effect of size parameter ($X$) and wavelength dependent complex refractive index ($m$) values. In our study, the wavelength dependent complex refractive indices of amorphous silicates and organic refractories are considered, which are taken from \cite{Sc1996} and \cite{Je1993}. The real part of complex refractive index ($n$) decreases  in case of silicates and increases in case of organic when wavelength ($\lambda$) changes from $0.45 \mu m$ to $0.65 \mu m$, whereas the imaginary part ($k$) increases for silicates and decreases for organic when $\lambda$ changes from $0.45 \mu m$ to $0.65 \mu m$.
Again, due to the high absorption efficiency of organic refractory at $\lambda$ = 0.45$\mu$m (larger $X$), ${S}_{11}(180^\circ$) is found to be lower as compared to the smaller $X$ associated with $\lambda$=0.65$\mu$m (since, $X_{0.45\mu m} > X_{0.65\mu m}$) (discussed in section 3).  ${S}_{11}(180^\circ$) is noticeably higher for the fluffy BA cluster as compared to that of the compact BAM2 cluster for the organic refractory. But the opposite trend is being well observed for the amorphous silicates where the BAM2 cluster possess higher ${S}_{11}(180^\circ$) than the fluffy BA cluster.
	
Most interesting results are observed in Fig.3 in the case of organic refractory where the polarization maximum significantly drops with the increase in wavelength for BAM1 and BAM2 clusters of 1024 monomers. This is due to the effect of size parameter and complex refractive index, as mentioned above. Since dark organic refractories show higher ${S}_{11}(180^\circ$) at lower size parameter, it may result a decrease in polarization maximum for compact aggregates at $\lambda$= 0.65$\mu$m. The effect of size parameter is studied in the next section. From the Fig. 3, the conclusion can be easily derived that the extent of depolarization enhances with the increase in wavelength as the porosity of the aggregate drops prominently which is one of the reason behind this abrupt feature. Further, the size parameter of the aggregates decreases with the increase of wavelength, so the effect of size parameter can not be ruled out here.
	
The negative polarization observed in comets at high scattering angles is well explained using the backscattering mechanism. Due to the interference of the multiply scattered light wave, the backscattering mechanism comes into play \citep{Mi2002}. At high scattering angle, the waves traveling in direct and opposite courses interfere constructively. This leads to an enhancement in intensity and in turn increases the amplitude of negative polarization mainly for the silicate composites. The rise in intensity and the negative polarization due to the interference of multiply scattered waves is mainly produced by the constituent monomers in the outer layers of the cluster where the radiation field is homogeneous \citep{Ti2004}.

It is worth to mention that the appearance of the NPB and the enhancement in the backscattering intensity are not always obtained simultaneously in modeling. The opposition effects are mainly defined by the interference of multiply scattered waves and the near-field effect \citep{Ti2004}. The near field effect stimulates the negative polarization branch, but the enhancement in intensity is almost absent, whereas both the intensity and the NPB are found to be pronounced in the interference of multiply scattered waves. A strong direct correlation between the increase in NPB and the enhancement in intensity in the backscattering direction for the amorphous silicate composites can be well noticed in this study. So, in our simulated work, the interference of multiply scattered waves plays a crucial role rather than the near-field interaction.

The phase function ratio in the backscattering region is given by ${S}_{11}(180^\circ$)/S11(150$^\circ$). For the aggregates of high absorbing organics refractory, we find a small rise in the phase function ratio. This ratio is about 1.3 for BA and 1.2 for BAM2 cluster of organic refractory near the backscattering direction. In contrast, the amorphous silicates show a strong backscattering rise where the ratio is about 1.8 for BA and 1.7 for BAM2 cluster respectively. These values are close to the comet phase function, where the rise from 150$^\circ$ to 180$^\circ$ scattering angle is about a factor of 2 \citep{Mi1982}. The strong backscattering of the aggregates of silicate compounds is the key reason behind the pronounced negative polarization branch observed in the comet.

Light scattering properties of different particles: glass beads, rough particles, polyhedral solids, dense aggregates and aggregates with porosity higher than 90\% were studied using the PROGRA$^2$ instrument \citep{Ha2007,Ha2009,Ha2011,Fr2011}. From the laboratory measurements of submicron sized aggregates of silica and carbon grains using the PROGRA$^2$ instrument, it has been found that in the same size range, the scattering angle $\theta_{max}$ is larger for transparent levitating silica particles than for absorbing carbon particles \citep{Ha2009,Ha2011}. The opposite is observed for deposited particles. We also try to compare our simulated results with experimental results keeping in mind that only qualitative comparison is possible. Since the PROGRA$^2$ experiments were conducted particularly for large aggregates having high porosity, one should not expect exact results from simulations  with moderately large aggregates. In our study, BCCA and BPCA/BA clusters are comparable with the sample aggregates considered in PROGRA$^2$ experiment. We have already observed from Figs. 2 and 3 that $\theta_{max}$ for BA cluster (as well as  BAM1 and BAM2 clusters) attains at a higher scattering angle for the amorphous silicate compositions as compared to the case of organic refractories, which are in good agreement with PROGRA$^2$ results.  Further, \cite{Fr2011} studied the light scattering properties of soot and carbon black aggregates using the PROGRA$^2$ instrument. They found that $P_{max}$ produced by the soot samples are obtained for scattering angles between $80^\circ$ and $90^\circ$ whereas for black-carbon aggregates, the value decreases from $80^\circ$ to $70^\circ$ when the primary particle size increases. In our simulated results (see Fig. 3), $\theta_{max}$ varies between $83^\circ$ and $94^\circ$ for organic refractories. This nature is also comparable with the experimental results.
	
In summary, the simulated curves for BA, BAM1 and BAM2 clusters of amorphous silicate show a pronounced rise in the maximum positive polarization peak with the increase in wavelength. Highly porous BA cluster shows a significant increment in positive polarization peak for the organic refractory with the increase in wavelength. But in the case of less fluffy aggregates (like BAM1 and BAM2), the polarization maximum is being found to dampen with the increase in wavelength for the organic refractories. This section also well explains the direct correlation between the enhancement in intensity in the backscattering direction and the production of stronger and wider negative branch of polarization. The high growth of the intensity in the backscattering direction for the silicate composites as compared to the minimal increment for the organic refractories is the key factor behind the stronger and deeper NPB observed at high scattering angle for the silicate composition.

\subsection{Effect of aggregate size parameter ($X$)}
\cite{Ki2006} and other investigators studied the effect of size parameters of aggregates (BCCA and BPCA) where porosities were considered to be very high ($\mathcal{P} \gtrsim 0.90$), but in our study, we explore the effect of aggregate size parameter in a wide range of porosity.
The study of the effect of size parameter ($X=2 \pi R/\lambda$) of the aggregate can be done either by changing the number of monomers ($N$) in the aggregate or by changing the radius of the constituent monomer (a$_m$). Since we have fixed $a_m$ at 0.1$\mu$m, the effect of $X$ can be realized when we change $N$.
Considering aggregates of $N = 256, 512, 1024$ monomers result in the effective size parameter of the aggregate BA ($X$= 16.8, 21.6, 27.3) and BAM2 ($X$=11.8, 15.3, 19.8) at $\lambda$=0.45$\mu$m. In Fig 4, we show the effect of size parameter of the aggregate ($X$) on phase function and polarization for BA (i and ii) and BAM2 (iii and iv) for amorphous silicates at $\lambda$ = 0.45$\mu m$ with $m = 1.689 + i~ 0.0031$.

 Fig. 4 demonstrates that the NPB does not show any significant dependence on the size parameter for the BA cluster that possess a value of about $-0.07$ at $169-172^\circ$ scattering angle. In contrast, the NPB displays a symmetric feature and it is found to be stronger and deeper for the compact BAM2 cluster with the increment in the size parameter ($-0.040, -0.045$ and $-0.057$ for $X$ = 11.8, 15.3, 19.8). Thus the increasing X by increasing the number of monomers enhance the NPB for compact BAM2 cluster but do not influence the NPB for fluffy BA cluster. The effect of size parameter on the positive  polarization maximum (P$_{max}$) and amplitude of negative polarization (P$_{min}$) are depicted in Table-2.
\begin{figure*}
\begin{center}
\includegraphics[width=130mm]{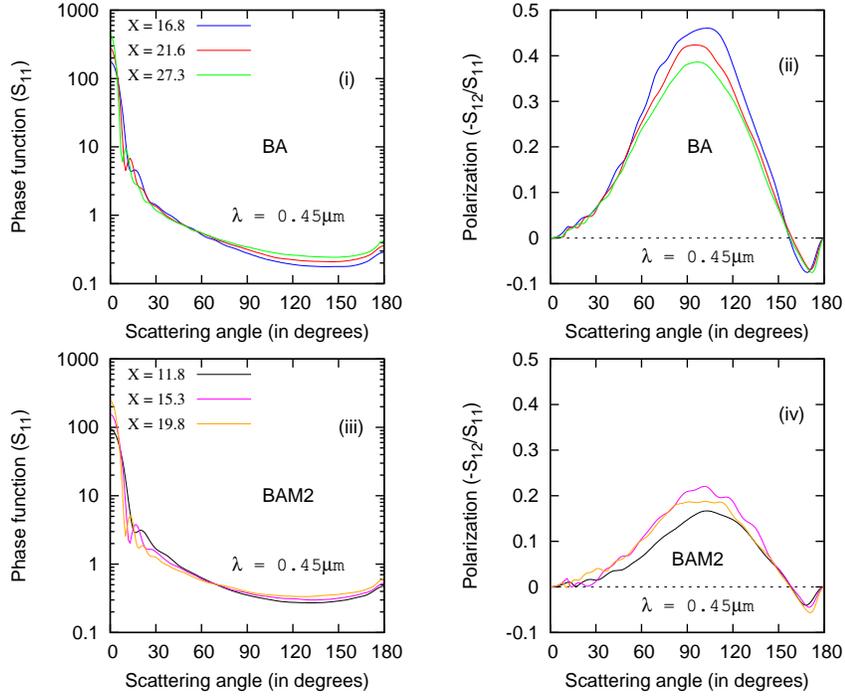}

  \caption{Effect of size parameter of the aggregate ($X$) on phase function and polarization for BA (i and ii) and BAM2 (iii and iv) for amorphous silicates at $\lambda$=0.45$\mu$m with $m = 1.689 + i~ 0.0031$. We consider aggregates of $N$ = 256, 512, 1024 monomers which result in the effective size parameter of the aggregate BA ($X$= 16.8, 21.6, 27.3) and BAM2 ($X$=11.8, 15.3, 19.8) at $\lambda$=0.45$\mu$m.}
\end{center}

\end{figure*}

\begin{figure*}
\begin{center}
\includegraphics[width=130mm]{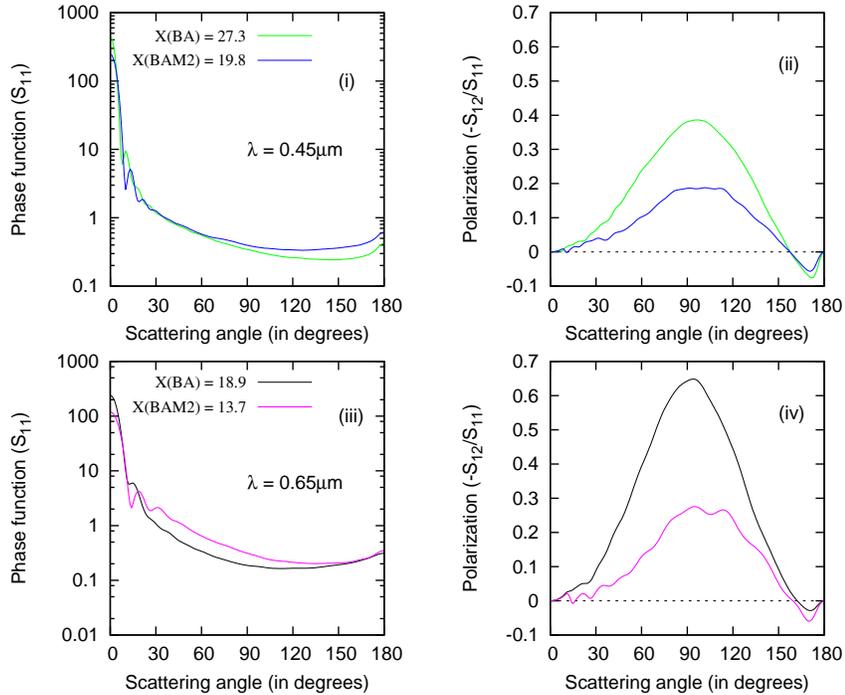}

  \caption{Effect of size parameter ($X$) on phase function and polarization for porous (BA) and compact (BAM2) clusters consisting of amorphous silicate composites with $m = 1.689 + i~ 0.0031$ at $\lambda$= 0.45$\mu$m and $m = 1.677 + i~ 0.0044$ at $\lambda$ = 0.65$\mu$m. Here the number of monomers (N = 1024) and size of the aggregate ($a_m$ = 0.1$\mu$m) are taken to be the same. The upper panel (i and ii) shows the results at $\lambda$=0.45$\mu$m and the lower panel (iii and iv) depicts the results at $\lambda$=0.65$\mu$m.}
\end{center}
\end{figure*}


\begin{table*}
\caption{Summary of results from Fig. 4 at $\lambda$ = 0.45$\mu m$. Types of aggregate, size parameter of the aggregate ($X$), scattering angle of polarization maximum ($\theta_{max}$), polarization maximum (P$_{max}$), inversion angle ($\theta_{inv}$),  scattering angle of polarization minimum ($\theta_{min}$), amplitude of negative polarization (P$_{min}$) and phase function at the exact back scattering direction ($S11(180^\circ)$).}
\begin{center}
\begin{tabular}{|c|c|c|c|c|c|c|c|}
\hline
   Aggregate & $X$=2$\pi$R/$\lambda$ & $\mathbf{\theta_{max}}$ & P$_{max}$ & $\mathbf{\theta_{inv}}$ & $\mathbf{\theta_{min}}$ & P$_{min}$ & ${S}_{11}(180^\circ$) \\
 \hline
   BA &  16.8 & 103$^\circ$ & 0.46 & 158$^\circ$ & 169$^\circ$ & --0.075 & 0.29 \\

    &  21.6 & 95$^\circ$ & 0.42 & 159$^\circ$ & 171$^\circ$ & --0.070 & 0.36 \\

    &  27.3 & 97$^\circ$ & 0.39 & 158$^\circ$ & 172$^\circ$ & --0.076 & 0.42 \\

    \hline
     BAM2 & 11.8 & 103$^\circ$ & 0.17 & 159$^\circ$ & 168$^\circ$ &--0.040 & 0.48 \\

    &  15.3 & 101$^\circ$ & 0.22 & 159$^\circ$ & 170$^\circ$ &--0.045 & 0.52 \\

    &  19.8 & 102$^\circ$ & 0.19 & 158$^\circ$ & 171$^\circ$ &--0.057 & 0.60 \\

   \hline
\end{tabular}
\end{center}
\end{table*}

	
Now to isolate the effect of size and number of the constituent monomers, we have considered two different types of aggregates (BA and BAM2) with the equal number of constituent monomers (N=1024) of same size 0.1$\mu$m which provide a large deviation in the size parameter ($X$) of the aggregate. This helps us to study the variation in the light scattering properties of aggregates with the significant change in the size parameter. In Fig. 5, we show the effect of size parameter ($X$) on phase function and polarization for BA and BAM2 clusters consisting of amorphous silicate composites with $m = 1.689 + i~ 0.0031$ (at $\lambda$= 0.45$\mu$m) and $m = 1.677 + i~ 0.0044$ (at $\lambda$ = 0.65$\mu$m). We have carried out our computation, keeping N fixed at 1024, with silicate aggregates of size parameter as high as $X$$\sim$27 (for BA cluster) and $X$$\sim$20 (for BAM2 cluster) at $\lambda$=0.45$\mu$m and $X$$\sim$19 (for BA cluster) and $X$$\sim$14 (for BAM2 cluster) at $\lambda$=0.65$\mu$m.

Fig. 5 depicts a higher degree of positive polarization for BA cluster as compared to BAM2 cluster in both the wavelengths. The simulated curve shows surprisingly higher negative polarization ($\sim -7\%$) for the BA cluster of highly porous nature as compared to that of the compact aggregate structure BAM2 ($\sim -6\%$) for the amorphous silicate composition at $\lambda$=0.45$\mu$m. The NPB obtained at $\lambda$=0.65$\mu$m for the BA cluster dampens ($\sim -3\%$) due to the significant decrease in the size parameter of the aggregate ($X$$\sim$19) and change in the complex refractive index from $\lambda$ = 0.45$\mu$m to 0.65$\mu$m. More interestingly, the amplitude and the width of the NPB at higher scattering angle for the compact BAM2 cluster do not show such strong dependence on the size parameter (X). Thus the NPB of fluffy aggregates is found to be prominent as compared to that of the compact BAM2 cluster when the size parameter ($X$) has been increased by changing the wavelength with monomer size and numbers are kept fixed.
	
It can be seen from Figs. 4 and 5 that ${S}_{11}(180^\circ$) for silicate composites is found to be higher at wavelength $\lambda$=0.45$\mu$m as compared to that at $\lambda$=0.65$\mu$m for both BA and BAM2 clusters which suggest that the photometric color is blue for silicate particles. This may be due to the larger size parameter ($X$) of the aggregate at $\lambda$=0.45$\mu$m which is the most significant factor affecting the light scattering properties of the aggregate. The higher size parameter results in the coverage of large scattering cross section of the bright amorphous silicates composites by the wavelength of radiation which enhance ${S}_{11}(180^\circ$). The silicate composites showed almost 10\% reflection for both the aggregates irrespective of their wide variation in the aggregation procedure of large porosity difference. The BA cluster of high size parameter ($X$$\sim$27) has shown a stronger enhancement in intensity in the backscattering direction as compared to that of the BAM2 cluster ($X$$\sim$20) at $\lambda$=0.45$\mu$m. This leads to the production of stronger and wider NPB for the highly porous BA cluster than the compact BAM2 cluster at high scattering angle.

\cite{Ha2007} presented linear polarization measurements of nine samples levitating particles (five samples of vapor-condensed magnesiosilica, one ferrosilica smoke, a mixture of magnesio–ferrosilica smokes, one mixture of ferrosilica with carbon and one mixture of magnesio–ferrosilica with carbon) using the PROGRA$^2$ instrument. They reported that $P_{max}$ is around 40\% (close to cometary values) and there is no wavelength dependence; a shallow negative polarization can be present at scattering angles greater than 160$^\circ$. In our study, $P_{max}$ and $\theta_{inv}$ for moderately large aggregates of amorphous silicate composition depicted in Table-2 for BA cluster at $\lambda$ = 0.45$\mu m$ are given by $\sim 40 \%$ and $\sim 159^\circ$, which are in good agreement with the PROGRA$^2$ experimental results for two samples of magnesiosilica (sample \#d and \#e)  which contain amorphous SiO$_2$ \citep{Ha2007}. Our findings show that the value of polarization maximum observed for large aggregates in PROGRA$^2$ experiment is also comparable with simulated results obtained for moderately large aggregates. The scattering angle corresponds to minimum polarization value for these two samples (sample \#d and \#e) are also found to be consistent with our simulated results. However, at higher wavelength $\lambda$ = 0.65$\mu m$, $P_{max}$ becomes 65\%, in case of BA cluster (see Fig. 5(iv)). This may be due to combined effect of size parameter ($X$) and wavelength dependent complex refractive index ($m$) values, as discussed earlier.

To summarize, in this study, we have varied the size parameter in two different ways (i) by changing the number of monomers and (ii) by changing the aggregate type to get a detailed picture of the effect of size parameter on the light scattering properties of dust aggregates. The effect of size parameter ($X$) plays a significant role in the optical properties of aggregated dust particles. The NPB of porous aggregate (BA) is found to be enhanced with the increase in the size parameter of the aggregate whereas the compact aggregate (BAM2) has shown a very weak dependence on $X$ when the monomer size and numbers are kept fixed. But the effect of size parameter on the NPB is found to be pronounced for compact BAM2 cluster in comparison to the fluffy BA cluster when the size parameter is being varied by changing the number of monomers in the aggregate.

\subsection{Effect of porosity}
To scrutinize how the porosity influences the polarization, particularly the negative branch of polarization for moderately large aggregates, we compute the polarization and phase function of the four different aggregates BCCA, BA, BAM1 and BAM2 having the porosity 0.98, 0.87, 0.74 and 0.64 for high albedo silicate particles.

Earlier \cite{Ti2004} and \cite{Pe2004} explained the effect of porosity on the polarization for small aggregates. But in this study, we have carried out our computation for considerably large aggregates ($R > \lambda$) of three different types having a wide variation in porosity to explore the behavior of light scattering properties of aggregates with the change in the porosity. \cite{Sh2009} reported that the apparent effects of monomer size on total cross sections are due to the effects of varying porosity ($\mathcal{P}$) or characteristic radius of aggregate ($R$). They commented that in previous studies \citep{Pe2000, Be2007}, variations of monomer size ($a_m$) were always coupled with changes in $\mathcal{P}$ and $R$, hence effects attributed to varying $a_m$ may be due to variations in $\mathcal{P}$ or $R$. \cite{Sh2009} also showed that the monomer size had little effect on phase function and polarization at constant $R$ and $\mathcal{P}$ (see Fig. 4 of \cite{Sh2009}).

We have performed our computation with the same characteristic radius (R$\sim$1$\mu$m, X$\sim$14) for all four aggregates of amorphous silicate composition at $\lambda$=0.45 $\mu$m with $N = 1024$. The size of the monomer ($a_m$) is considered to be different of about 0.03, 0.05, 0.06 and 0.07$\mu$m for BCCA, BA, BAM1, and BAM2 clusters to get the characteristic radius of the aggregate (R)$\sim$1$\mu$m. Fig. 6 displays that the most porous BCCA cluster shows highest polarization ($\sim 93 \%$) in the positive branch, but it does not show any sign of negative polarization at this size.  BA cluster shows  maximum polarization of $\sim 79 \%$ , but the negative polarization branch is found to be very small ($-1\%$). The moderately porous BAM1 cluster produces a comparatively lower positive polarization peak ($\sim 53 \%$) than the highly porous clusters (BCCA and BA), but it produces a considerably higher negative polarization ($-4\%$) at high scattering angle range. The highly dense BAM2 cluster with a very compact structure produces much deeper and wider NPB ($-6\%$) than the other three types of aggregates, but the peak of the positive polarization becomes very low ($\sim 26 \%$).

Fig. 7 demonstrates the extent of dependence of $P_{min}$, $S_{11}(180^\circ)$ (or geometric albedo) and $P_{max}$ on the porosity ($\mathcal{P}$) for four considered aggregates. We have observed that they are correlated linearly with $\mathcal{P}$ when the porosity changes from 0.64 to 0.98. Compact BAM2 structure shows the highest degree of negative polarization, highest $S11$(180$^\circ$) and lowest $P_{max}$ as compared to other three clusters when the characteristic radius ($R$) of all clusters are considered to be the same. We find that phase function at the exact back scattering direction (${S}_{11}(180^\circ$)) and the amplitude of negative polarization ($P_{min}$) decrease linearly with the increase in $\mathcal{P}$ whereas the polarization maximum (P$_{max}$)  increases linearly with the increase in $\mathcal{P}$.

\begin{figure*}
\begin{center}
\includegraphics[width=120mm]{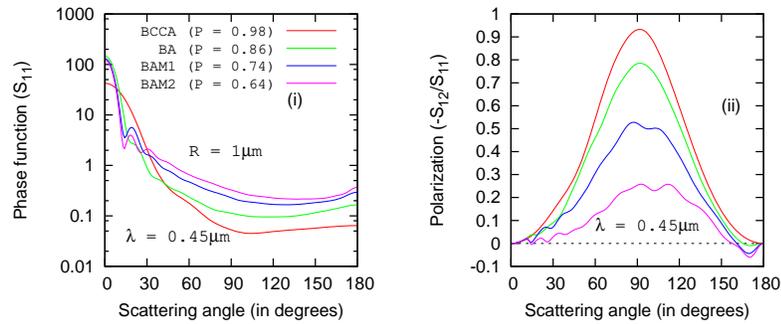}

  \vspace{-3.5cm}
  \caption{Effect of porosity on (i) Phase function and (ii) polarization for four clusters (BCCA, BA, BAM1 and BAM2) at $\lambda$=0.45$\mu$m. The characteristic radius (R) of each aggregate is 1$\mu$m, where N = 1024.}
\end{center}

\end{figure*}


\begin{figure*}
\begin{center}
\includegraphics[width=110mm]{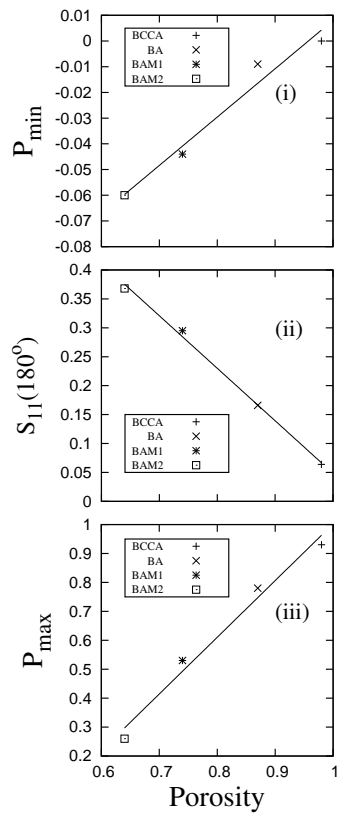}

  \caption{Variation of $P_{min}$, $S_{11}(180^\circ)$ and $P_{max}$ with the porosity ($\mathcal{P}$) of aggregates at $\lambda$=0.45$\mu$m. The values are taken from Fig.6. The solid lines in $(i)$, $(ii)$ and $(iii)$ indicate the best fit lines. The characteristic radius ($R$) of each aggregate is 1$\mu$m, where N = 1024.}
\end{center}

\end{figure*}

For porous BA cluster, the negative branch in the backscattering direction is very low due to the significant inter-monomer spacing which results in the weak wave interference and near-field effects within the aggregates. In contrast, the light scattered by a compact BAM2 cluster is found to be depolarized, and a strong negative polarization is well noticed at a high scattering angle, probably, indicating a stronger electromagnetic interaction between the constituent monomers present in the aggregate. The depolarization phenomena in the aggregate are due to the occurrence of cross-polarization in contrast to that for a spherical grain. The presence of depolarization is an indicator of multiple scattering which occurs due to the intracluster scattering which gives an interesting insight of the internal structure of the aggregate \citep{Pe2004}. Multiple scattering is of crucial importance in the vicinity of a comet nucleus where dust particles are released from their parent bodies as well as on the surface of their parent bodies \citep{Ki2016}.

Fig. 6 shows that the polarization maximum shifts towards the higher scattering angle as the porosity of the aggregate decreases. In contrast, when the porosity of the aggregate decreases inversion angle of the polarization curve shift towards lower scattering angle.
	
It is important to mention that a resonance like structure can be well noticed in the polarization curve of considered compact BAM1 and BAM2 clusters typical for solid particles of this size. Since in BAM2 cluster, the monomers are very densely packed, the incident light interacts with the aggregate as a whole which results in such ripple-like structure. The similar nature of resonance structure for the densely packed aggregates was reported by \cite{Ti2004} and \cite{Ok2008}. The coupling between the monomers comes into play only when the size and the intermediate spacing of the monomers are less than the wavelength of radiation. Thus the coupling will be higher for the compact aggregates as compared to the porous one as a single wavelength covered a large number of monomers which lead to a higher degree of depolarization of the scattered radiation. It has also been found that the extent of depolarization effect becomes more significant for the aggregates of larger size than the wavelength of radiation \citep{Ti2004}.
				
The left panel of Fig. 6 displays the photometric feature of the four considered dust aggregates of different porosity at $\lambda$=0.45$\mu$m. It can be seen from the figure that ${S}_{11}(180^\circ$) is highest for BAM2 particles and lowest for BA particles. Thus the phase function is correlated with the porosity of the aggregate, i.e., higher porosity has lower ${S}_{11}(180^\circ$), and the lower porosity aggregates possess higher ${S}_{11}(180^\circ$) in case of silicate composition. The enhancement in intensity in the backscattering direction is higher for less porous aggregates which result in the production of stronger and wider negative polarization branch for silicate composites.
		
We can summarize that the effect of porosity is also of crucial importance in the study of light scattering properties of aggregates. The geometric albedo or  ${S}_{11}(180^\circ$), $P_{max}$ and $P_{min}$ show a direct correlation with the porosity of the aggregates. The amplitude of negative polarization ($P_{min}$) and phase function $({S}_{11}(180^\circ))$ are found to be enhanced with the decrease in porosity of the aggregates whereas $P_{max}$ is found to be decreased with the decrease in porosity from 0.98 to 0.64.

\subsection{Effect of mixing}

Figs. 8 and 9 depict the photopolarimetric feature of porous BA and most compact BAM2 clusters. The plots are being generated by a suitable mixing of organic and silicate composites. The NPB is found to be stronger with the decrease in proportion of organic composites ($-0.021$, $-0.035$ and $-0.045$) for 90\%, 80\% and 70\% of organics for BA cluster at $\lambda$=0.45$\mu$m and ($-0.010$, $-0.013$ and $-0.016$) at $\lambda$=0.65$\mu$m. The enhancement in the NPB with the decrease in organic composites in the mixing is also well observed for BAM2 cluster ($-0.019$, $-0.031$ and $-0.039$) for 90\%, 80\% and 70\% of organics at $\lambda$=0.45$\mu$m and ($-0.015$, $-0.027$ and $-0.035$) at $\lambda$=0.65$\mu$m. In Fig. 10, the polarimetric color of BA and BAM2 clusters with the variation in scattering angle is plotted. The figure shows that the polarimetric color (P$_{0.65}$$-$P$_{0.45}$) is positive for the BA cluster and the value decreases with the increase in the proportion of organic refractory in the mixing. But a completely reverse trend is being monitored in case of compact BAM2 cluster where the polarimetric color is found to be negative beyond an intermediate scattering angle irrespective of the change in mixing proportions. At this stage, we should keep in mind that this dust model is not appropriate for comet dust modeling as the size of comet dust may range from several tens to hundreds of microns which were detected in both the space mission and laboratory diagnosis of comet Wild-2 dust (\cite{Ko2007} and references therein) and recent Rosetta mission on Comet 67P/Churyumov-Gerasimenko \citep{Sc2015}. Since computational restrictions stop us from trying this for very large aggregates, so we can conclude that this dust model could deal with a small aggregated particles or in the sense of moderately large aggregates as compared to the wavelength of radiation.

\begin{figure*}
\begin{center}
\hspace{2.0cm}
\includegraphics[width=110mm]{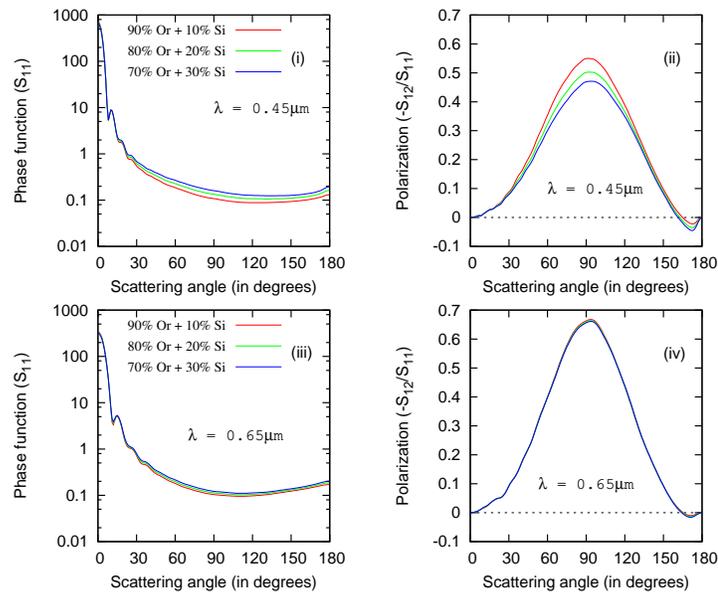}

	\vspace{1cm}
  \caption{Photopolarimetric characteristics of BA for different mixing ratios of amorphous silicates and organic refractory composites. The characteristic radius (R$\sim$1.95$\mu$m) and porosity ($\mathcal{P}$$\sim$0.87) of BA are taken to be the same at the two wavelengths 0.45$\mu$m and 0.65$\mu$m. Phase function and polarization of the aggregate obtained from this work at $\lambda$=0.45$\mu$m (i and ii) and 0.65$\mu$m (iii and iv) are shown.}

\end{center}

\end{figure*}


\begin{figure*}
\begin{center}
\hspace{2.0cm}
\includegraphics[width=110mm]{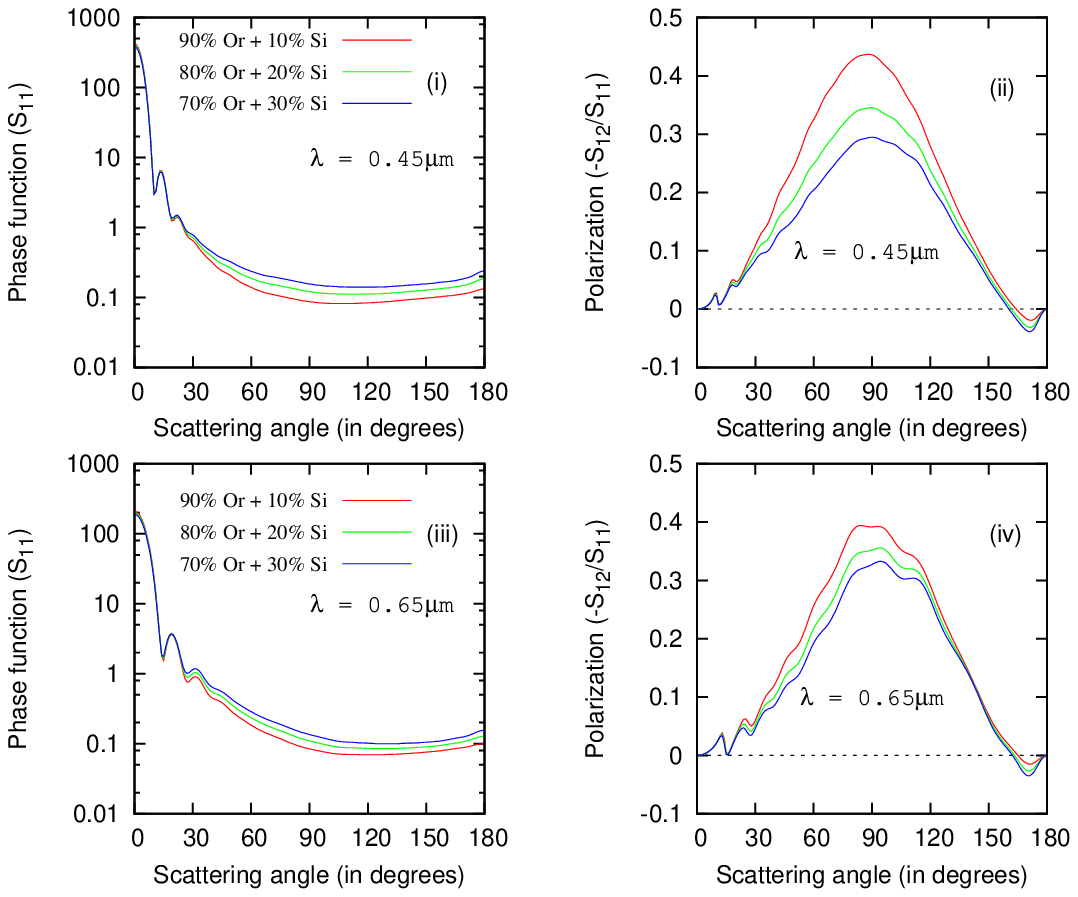}

	\vspace{1cm}
  \caption{Photopolarimetric characteristics of BAM2 for different mixing ratios of amorphous silicates and organic refractory composites. The characteristic radius (R$\sim$1.4$\mu$m) and porosity ($\mathcal{P}$$\sim$0.64) of BAM2 are taken to be the same at the two wavelengths 0.45$\mu$m and 0.65$\mu$m. Phase function and polarization of the aggregate obtained from this work at $\lambda$=0.45$\mu$m (i and ii) and 0.65$\mu$m (iii and iv) are shown.}
\end{center}

\end{figure*}

\begin{figure*}
\begin{center}
\includegraphics[width=120mm]{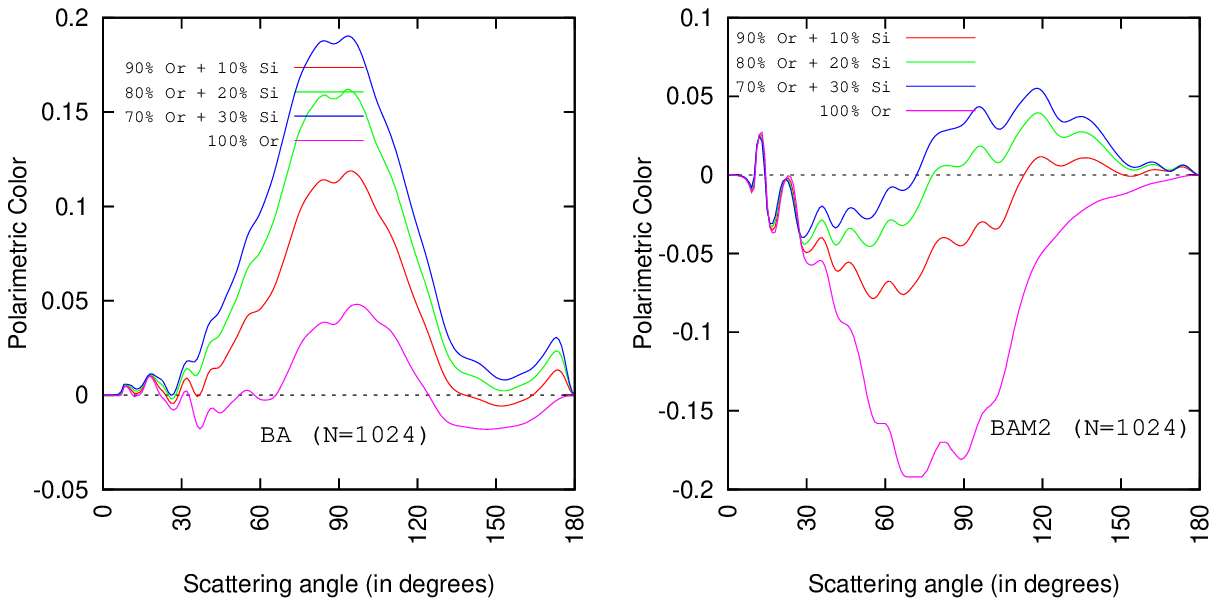}

  \vspace{-0.8cm}
 \caption{Polarimetric color $(P_{0.65} - P_{0.45})$ for BA (left panel) and BAM2 (right panel) clusters of different mixing ratios of amorphous silicate and organic refractory. The result with 100\% organic composition is also drawn for reference.}
\end{center}

\end{figure*}


\section{Conclusions}
\begin{enumerate}
 \item[(i)] The effect of aggregate size parameter ($X$) on the light scattering properties of aggregates (BA and BAM2) having different porosities is explored in this study. It is found that the negative polarization branch (NPB) for compact BAM2 cluster enhances more as compared to NPB for BA cluster if $X$ is increased by increasing $N$. But, an opposite trend is being noticed if $X$ is increased by changing the wavelength, N being kept constant, where enhancement of NPB for highly porous BA cluster is found to be more prominent as compared to less porous BAM2 cluster.

\item[(ii)] We have found from our study that the positive polarization maximum ($P_{max}$), the amplitude of the negative polarization ($P_{min}$) and phase function at the exact back scattering direction (${S}_{11}(180^\circ$)) are correlated with the porosity of aggregates. Compact aggregates show deeper negative polarization as compared to porous aggregates when the characteristic radius ($R$) of the aggregates is considered to be the same. Further lower porosity aggregates show higher ${S}_{11}(180^\circ)$ and higher porosity aggregates show comparatively lower ${S}_{11}(180^\circ)$.

\item[(iii)] We have also investigated the correlation of P$_{max}$, P$_{min}$ and $S_{11}(180^\circ)$ with porosity ($\mathcal{P}$) for silicate composites.  It is found from our study that when $\mathcal{P}$ is increased in a range from 0.64 to 0.98, both ${S}_{11}(180^\circ$) and $P_{min}$ decrease linearly, whereas P$_{max}$ increases linearly.

\item[(iv)] We have observed that the porosity of aggregates plays a crucial role in determining the polarimetric color for high absorbing organic
    refractories. The compact aggregates like BAM1 and BAM2 show the negative polarimetric color whereas BA cluster shows almost positive polarimetric color at all values of scattering angle.
\end{enumerate}

\section*{Acknowledgements}
We acknowledge Daniel Mackowski and Michael  Mishchenko, who made their Multi-sphere T-matrix code publicly available. We also acknowledge Bruce T. Draine who made BA, BAM1 and BAM2 clusters publicly available in his website. The anonymous reviewer of this paper is highly acknowledged for useful comments and suggestions. This work is supported by the Science and Engineering Research Board (SERB), a statutory body under Department of Science and Technology (DST), Government of India, under Fast Track scheme for Young Scientist (SR/FTP/PS-092/2011). The author P. Deb Roy also wants to acknowledge DST INSPIRE scheme for the fellowship. We also acknowledge HPC centre of NIT Silchar in collaboration with C-DAC Pune, where some part of computations were performed.





\begin{thebibliography}{99}

\bibitem [\protect\citeauthoryear{Bertini et al.}{2007}]{Be2007}
Bertini I., Thomas N., Barbieri C., 2007, A\&A, 461, 351.

\bibitem [\protect\citeauthoryear{Bockel et al.}{2002}]{Bo2002}
Bocke´ ee - Morvan D., Gautier D., Hersant F., Hur´ e J. M., \& Robert F., 2002, A\&A 384, 1107.

\bibitem [\protect\citeauthoryear{Brownlee et al.}{1980}]{Br1980}
Brownlee D. E., Pilachowski L., Olszewski E., Hodge P. W., 1980, Analysis of interplanetary dust collections. In: Halliday I., McIntosh B. A., editors. Solid particles in the solar system. Dordrecht: D. Reidel, p. 333-342.

\bibitem [\protect\citeauthoryear{Capaccioni et al.}{2015}]{Ca2015}
Capaccioni F., Coradini A., Filacchione G., Erard S., Arnold G., et al., 2015, Science, 347, 6220

\bibitem [\protect\citeauthoryear{Das et al.}{2004}]{Da2004}
Das H. S., Sen A. K., Kaul C. L., 2004, A\&A, 423, 373.

\bibitem [\protect\citeauthoryear{Das et al.}{2006}]{Da2006}
Das H. S., Sen A. K., 2006, A\&A, 459, 271

\bibitem [\protect\citeauthoryear{Das et al.}{2008}]{Da2008}
Das H. S., Das S. R., Sen A. K., 2008, MNRAS, 390, 1190.

\bibitem [\protect\citeauthoryear{Das et al.}{2011}]{Da2011}
Das H. S., Paul D., Suklabaidya A., Sen A. K., 2011, MNRAS, 416, 94.

\bibitem [\protect\citeauthoryear{Dollfus et al.}{1988}]{Do1988}
Dollfus A., Bastien P., Le Borgne J. F., 1988, A\&A, 206, 348.

\bibitem [\protect\citeauthoryear{Eaton et al.}{1992}]{Ea1992}
Eaton N., Scarrott S. M., Gledhill T. M., 1992,  MNRAS, 258, 384.

\bibitem [\protect\citeauthoryear{Francis et al.}{2011}]{Fr2011}
Francis, M., J.-B. Renard, E. Hadamcik, et al., 2011, J. Quant. Spectrosc. Radiat. Transfer, 112, 1766

\bibitem [\protect\citeauthoryear{Goesmann et al.}{2015}]{Go2015}
Goesmann F., Rosenbauer H., Bredeh{\"o}ft J. H., Cabane M., Ehrenfreund P., et al., 2015, Science, 349, 6247

\bibitem [\protect\citeauthoryear{Hadamcik et al.}{2007}]{Ha2007}
Hadamcik E., Renard J.-B., Rietmeijer F. J. M. , Levasseur-Regourd A. C., Hill H. G. M., Karner J. M., Nuth J. A., 2007, Icarus, 190, 660.

\bibitem [\protect\citeauthoryear{Hadamcik et al.}{2009}]{Ha2009}
Hadamcik E., Renard J.-B., Levasseur-Regourd A. C., Lasue J., Alcouffe G., Francis M., 2009, JQSRT, 110, 1755.

\bibitem [\protect\citeauthoryear{Hadamcik et al.}{2011}]{Ha2011}
Hadamcik E., Renard J.-B., Levasseur-Regourd A. C., Lasue J., 2011, Laboratory measurements of light scattered by clouds and layers of solid particles using an imaging technique. In Polarimetric Detection, Characterization and Remote Sensing, pp. 137-176. Springer Netherlands

\bibitem [\protect\citeauthoryear{Hanner et al.}{2004}]{Ha2004}
Hanner M., Bradley J., 2004, Physical properties of cometary dust from light scattering and thermal emission. In: Festou, M., Keller, U., Weave, H. (Eds.), Comets II. University of Arizona Press, Tuscon., 555-564.

\bibitem [\protect\citeauthoryear{Hayward et al.}{2000}]{Ha2000}
Hayward T. L., Hanner M. S., Sekanina Z., 2000, ApJ, 538, 428.

\bibitem [\protect\citeauthoryear{Jenniskens et al.}{1993}]{Je1993}
Jenniskens P., 1993, A\&A, 274, 653.

\bibitem [\protect\citeauthoryear{Jessberger et al.}{1988}]{Je1988}
Jessberger E. K., Christoforidis A., Kissel J., 1988, Nature, 332, 691.

\bibitem [\protect\citeauthoryear{Jessberger et al.}{1999}]{Je1999}
Jessberger E. K., 1999, SSR, 90, 91.

\bibitem [\protect\citeauthoryear{Kimura et al.}{2006}]{Ki2006}
Kimura H., Kolokolova L., Mann I., 2006, A\&A, 449, 1243.

\bibitem [\protect\citeauthoryear{Kimura et al.}{2016}]{Ki2016}
Kimura H., Kolokolova L., Li A., Lebreton J., 2016, Light Scattering Reviews.

\bibitem [\protect\citeauthoryear{Kolokolova et al.}{2004}]{Ko2004}
Kolokolova L., Hanner M. S., Levasseur-Regourd A. C., Gustafson B. A. S., 2004, Physical properties of cometary dust from light scattering and thermal emission in Comets. In: Festou M, Keller U, Weaver H. (Eds). Arizona University Press., 577-605

\bibitem [\protect\citeauthoryear{Kolokolova et al.}{2007}]{Ko2007}
Kolokolova L., Kimura H., Kiselev N., Rosenbush V., 2007, A\&A, 463, 1189.

\bibitem [\protect\citeauthoryear{Kolokolova et al.}{2015}]{Ko2015}
Kolokolova L., Das H. S., Dubovik O., Lapyonok T., Yang P., 2015, PSS, 116, 30.

\bibitem [\protect\citeauthoryear{Levasseur et al.}{1996}]{Le1996}
Levasseur-Regourd A., C., Hadamcik E., Renard J. B., 1996, A\&A, 1996, 313, 327

\bibitem [\protect\citeauthoryear{Mackowski \& Mishchenko}{2013}]{Ma2013}
Mackowski D. W., Mishchenko M. I. A multiple sphere T-matrix FORTRAN code for use on parallel computer clusters Version 3.0. 2013

\bibitem [\protect\citeauthoryear{Markkanen et al.}{2015}]{Ma2015}
Markkanen J., Penttilla A., Peltoniemi J., Muinonen K., 2015, PSS, 118, 164.

\bibitem [\protect\citeauthoryear{Mazarbhuiya \& Das}{2017}]{Maz2017}
Mazarbhuiya, A. M. \& Das, H. S., 2017, Astrophys Space Sci, 362, 161

\bibitem [\protect\citeauthoryear{Millis et al.}{1982}]{Mi1982}
Millis R. L., A'Hearn M. F., Thompson D. T., 1982, AJ, 87, 1310.

\bibitem [\protect\citeauthoryear{Mishchenko et al.}{2002}]{Mi2002}
Mishchenko M., Tishkovets V., Litvinov P., 2002, Exact results of the vector theory of coherent backscattering from discrete random media: an overview. In: Videen G., Kocifaj M., editors. Optics of cosmic dust. Dordrecht: Kluwer, p. 239-260.

\bibitem [\protect\citeauthoryear{Muinonen et al.}{1991}]{Mu1991}
Muinonen K. O., Sihvola A. H., Lindell IV., and Lumme K. A., 1991, 8, 477.

\bibitem [\protect\citeauthoryear{Okada et al.}{2008}]{Ok2008}
Okada Y., Mann I., Mukai T., Kohler M., 2008, JQSRT, 109, 2613.

\bibitem [\protect\citeauthoryear{Petrova et al.}{2000}]{Pe2000}
Petrova E. V., Jockers K., Kiselev N. N., 2000, Icarus, 148, 526.

\bibitem [\protect\citeauthoryear{Petrova et al.}{2004}]{Pe2004}
Petrova E. V., Tishkovets V. P., Jockers K., 2004, SSR, 38, 309.

\bibitem [\protect\citeauthoryear{Renard et al.}{1996}]{Re1996}
Renard J.-B., Hadamcik E., Levasseur-Regourd A.-C., 1996, Astron. Astrophys., 316, 263.

\bibitem [\protect\citeauthoryear{Rietmeijer et al.}{2008}]{Ri2008}
Rietmeijer F. J., 2008, Understanding the comet Wild 2 mineralogy in samples from the Stardust mission, 23(02), 74.

\bibitem [\protect\citeauthoryear{Sandford et al.}{2006}]{Sa2006}
Sandford S. A. et al., 2006, Science, 314, 1720.

\bibitem [\protect\citeauthoryear{Schulz et al.}{2015}]{Sc2015}
Schulz R., Hilchenbach M., Langevin Y et al., 2015, Nature, 518, 216.

\bibitem [\protect\citeauthoryear{Scott \& Duley}{1996}]{Sc1996}
Scott A., Duley W. W., 1996, ApJ Suppl. Ser., 105, 401.

\bibitem [\protect\citeauthoryear{Shen et al.}{2008}]{Sh2008}
Shen Y., Draine B. T., Johnson E. T., 2008, ApJ, 689, 260.

\bibitem [\protect\citeauthoryear{Shen et al.}{2009}]{Sh2009}
Shen Y., Draine B. T., Johnson E. T., 2009, ApJ, 696, 2126.

\bibitem [\protect\citeauthoryear{Tishkovets et al.}{2004}]{Ti2004}
Tishkovets V. P., Petrova E. V., Jockers K., 2004, JQSRT, 86, 241

\bibitem [\protect\citeauthoryear{Wooden et al.}{1999}]{Wo1999}
Wooden D. H., Harker D. E., Woodward C. E., Butner H. M., Koike C., Witteborn F. C., McMurtry C. W., 1999, ApJ, 517, 1034.

\bibitem [\protect\citeauthoryear{Zolenky et al.}{2006}]{Zo2006}
Zolensky M. E. et al., 2006, Science, 314, 1735.



\end{thebibliography}



\end{document}